\documentclass[a4paper,12pt,twoside]{article}

\usepackage[dvips]{graphicx}
\usepackage{geometry}
\usepackage[centertags]{amsmath}
\usepackage{amsfonts}
\usepackage{amssymb}
\usepackage{amsthm}
\usepackage{latexsym}
\usepackage{mathrsfs}
\usepackage{fancyhdr}

\newcommand{\uD}{\ensuremath{\mathrm{d}}}
\newcommand{\uE}{\ensuremath{\mathrm{e}}}
\newcommand{\uI}{\ensuremath{\mathrm{i}}}

\renewcommand{\vec}[1]{\ensuremath{\boldsymbol{\mathrm{#1}}}}
\newcommand{\Cdot}{\ensuremath{\boldsymbol{\cdot}}}

\newcommand{\commute}[2]{\ensuremath{\left[{#1}\!%
\mathrel{\vphantom{{#1}},\vphantom{{#2}}%
\kern-\nulldelimiterspace}\!{#2}\right]}}
\newcommand{\anticommute}[2]{\ensuremath{\left\{{#1}\!%
\mathrel{\vphantom{{#1}},\vphantom{{#2}}%
\kern-\nulldelimiterspace}\!{#2}\right\}}}

\newcommand{\ArXivNo}{hep-th/0411273}

\newcommand{\XXXSize}{{\fontsize{12}{12}\selectfont\fbox{\textbf{\ArXivNo}}}}
\newcommand{\XXXTitle}{\hfill\XXXSize\newline\vskip 0.4cm}

\begin{document}
\pagestyle{fancy}

\sloppy

\title{\XXXTitle\textbf{Field Theory Methods in Classical Dynamics}%
\thanks{Work supported by the Royal Golden Jubilee Ph.D. Program.}~
\thanks{Published in \textit{International Journal of Theoretical Physics},
Vol.~\textbf{41}, No.~7, (2002) pp.~1327--1337. \ %
[\url{http://dx.doi.org/10.1023/A:1019607117138}] %
}}

\author{\textsc{E.~B.~Manoukian}\thanks{E-mail: \texttt{edouard@ccs.sut.ac.th}} \ and
\ \textsc{N.~Yongram} \\
{School of Physics, \ Suranaree University of Technology} \\
\ Nakhon Ratchasima, 30000, Thailand }
\date{} \maketitle

\begin{abstract}
A Dirac picture perturbation theory is developed for the time
evolution operator in classical dynamics in the spirit of the
Schwinger-Feynman-Dyson perturbation expansion and detailed rules
are derived for computations\@. \     Complexification formalisms
are given for the time evolution operator suitable for phase space
analyses, and then extended to a two-dimensional setting for a
study of the geometrical Berry phase as an example\@. \   Finally
a direct integration of Hamilton's equations is shown to lead
naturally to a path integral expression, as a resolution of the
identity, as applied to arbitrary functions of
generalized coordinates and momenta\@. \\

\noindent KEY WORDS: quantum field theory; time evolution in
classical dynamics; phase space complexification; Berry phase;
path integrals\@. \\
\end{abstract}

\section{Introduction}\label{Sec1}
With the rapid progress of field theory in describing the basic
interactions occurring in nature, it is becoming more and more
important to extend the powerful techniques of field theory in
describing physics, in general, in a more unified language\@. \
Such a unification has always been the goal of physics research
over the years\@. \     The present work is aimed in describing
classical dynamics in the spirit of field theory methods\@. \ Some
earlier papers on this subject are given in Abrikosov
(1993)~\cite{Abrikosov_1993}, Gozzi~\emph{et~al.}
(1989)~\cite{Gozzi_1989}, Schwartz (1976)~\cite{Schwartz_1977},
and Wetterich (1997)~\cite{Wetterich_1997}, which are, however,
only tangentially related to our present investigations\@. \ We
develop a Dirac (interaction) picture perturbation theory in the
spirit of the Schwinger-Feynman-Dyson perturbation theory to all
orders in a coupling parameter and detailed \emph{rules} are
derived for computations starting from the time evolution of any
function of the generalized coordinates and momenta\@. \
Complexifications of the time evolution are developed suitable for
phase space analyses and then extended to a two-dimensional
setting to describe the so-called geometrical Berry phase (Berry,
1984~\cite{Berry_1984}; Shapere and Wilczek,
1989~\cite{Shapere_1989}) as an example\@. \      Finally, since
the classical limit of the path integral, starting from the
quantum regime, reduces to just a phase factor involving the
classical action, such a limit (Abrikosov,
1993~\cite{Abrikosov_1993}; Gozzi~\emph{et~al.},
1989~\cite{Gozzi_1989}) is not of very practical value, as it
stands, for actual computations\@. \     Instead, we develop a
path integral expression by direct integration of Hamilton's
equations, as a resolution of the identity, that may be applied to
any function of the generalized coordinates and momenta, in the
same spirit of developing the resolution of the identity of a
self-adjoint operator, in quantum physics, that may be applied to
any vector in the underlying Hilbert space\@. \    The
perturbation expansion, with the derived rules, is given in
Section~\ref{Sec2}, while the complexification formalisms are
developed in Section~\ref{Sec3}\@. \     Section~\ref{Sec4} deals
with the path integral expression and its consistency with the
Poisson-bracket solution\@. \\

\renewcommand{\theequation}{\thesection.\arabic{equation}}
\numberwithin{equation}{section}
\section{Rules for Computations and the Dirac Picture
Perturbation Theory}\label{Sec2}

The time derivative of an arbitrary function $f[q(t),p(t)]$ of
generalized coordinates and generalized momenta, via the
Poisson-bracket formalism, reads
\begin{align}
  \frac{\uD}{\uD{}t}f[q(t),p(t)] &= \left[
  \frac{\partial{}H(t)}{\partial{}p(t)}\frac{\partial}{\partial{}q(t)}
  -\frac{\partial{}H(t)}{\partial{}q(t)}\frac{\partial}{\partial{}p(t)}\right]
  f[q(t),p(t)] \nonumber \\
  &\equiv{} f_{1}[q(t),p(t)]
  \label{Eqn2.1}
\end{align}
where $H[q(t),p(t)]\equiv{}H(t)$ is the Hamiltonian, constructed
out of the variables $q(t)$ and $p(t)$, assumed with no explicit
time dependence\@. \     Similarly,
\begin{align}
  \left(\frac{\uD}{\uD{}t}\right)^{\!n}\!f[q(t),p(t)] &\equiv{} f_{n}[q(t),p(t)]
  \nonumber \\
  &= \left[\frac{\partial{}H(t)}{\partial{}p(t)}\frac{\partial}{\partial{}q(t)}
  -\frac{\partial{}H(t)}{\partial{}q(t)}\frac{\partial}{\partial{}p(t)}\right]^{n}
  f[q(t),p(t)]
  \label{Eqn2.2}
\end{align}
leading to the familiar time evolution
\begin{equation}\label{Eqn2.3}
  f[q(t),p(t)] = \uE^{t\widehat{O}}f[q,p]
\end{equation}
where
\begin{equation}\label{Eqn2.4}
  \widehat{O} = \left[\frac{\partial{}H}{\partial{}p}\frac{\partial}{\partial{}q}
  -\frac{\partial{}H}{\partial{}q}\frac{\partial}{\partial{}p}\right]
\end{equation}
and $q=q(0)$, $p=p(0)$, $H=H(0)$\@. \\

Upon using the integral identity
\begin{equation}\label{Eqn2.5}
  \uE^{t[\widehat{A}+\lambda\widehat{B}]}\uE^{-t\widehat{A}} = 1
  +\lambda\int_{0}^{t}\!\!\uD{}\tau\:
  \uE^{\tau[\widehat{A}+\lambda\widehat{B}]}\widehat{B}\uE^{-\tau\widehat{A}}
\end{equation}
for two operators $\widehat{A}$ and $\widehat{B}$, and setting
\begin{align}
  H &= H_{1}+\lambda{}H_{2} \label{Eqn2.6} \\
  \widehat{A} &= \left[\frac{\partial{}H_{1}}{\partial{}p}\frac{\partial}{\partial{}q}
  -\frac{\partial{}H_{1}}{\partial{}q}\frac{\partial}{\partial{}p}\right]
  \label{Eqn2.7} \\
  \widehat{B} &= \left[\frac{\partial{}H_{2}}{\partial{}p}\frac{\partial}{\partial{}q}
  -\frac{\partial{}H_{2}}{\partial{}q}\frac{\partial}{\partial{}p}\right]
  \label{Eqn2.8}
\end{align}
one obtains, upon iteration of (\ref{Eqn2.5}), the expression
\begin{equation}\label{Eqn2.9}
  f[q(t),p(t)] = \sum\limits_{n=0}^{\infty}\lambda^{n}\int_{0}^{t}\!\!\uD{}t_{1}
  \int_{0}^{t_{1}}\!\!\uD{}t_{2}\cdots\int_{0}^{t_{n-1}}\!\!\uD{}t_{n}\:
  \widehat{B}(t_{n})\cdots{}\widehat{B}(t_{1})\,\uE^{t\widehat{A}}f[q,p]
\end{equation}
where
\begin{equation}\label{Eqn2.10}
  \widehat{B}(t) = \uE^{t\widehat{A}}\widehat{B}\uE^{-t\widehat{A}}
\end{equation}
with $\widehat{B}(t)$ independent of the parameter $\lambda$\@. \\

In particular, for
\begin{align}
  H_{1} &= \frac{p^{2}}{2m},\qquad{} H_{2}=V(q)
  \label{Eqn2.11} \\
  \widehat{B}(t) &= \widetilde{F}\!\left[q+\frac{t}{m}p\right]\left[
  \frac{\partial}{\partial{}p}-\frac{t}{m}\frac{\partial}{\partial{}q}\right]
  \label{Eqn2.12} \\
  \widetilde{F}[q] &= -V'(q)
  \label{Eqn2.13} \\
  f[q,p] &= q
  \label{Eqn2.14}
\end{align}
then
\begin{equation}\label{Eqn2.15}
  \uE^{t\widehat{A}}f[q,p] = q+\frac{t}{m}p
\end{equation}
and we obtain
\begin{equation}\label{Eqn2.16}
  q(t) = \sum\limits_{n=0}^{\infty}\lambda^{n}\int_{0}^{t}\!\!\uD{}t_{1}
  \int_{0}^{t_{1}}\!\!\uD{}t_{2}\cdots\int_{0}^{t_{n-1}}\!\!\uD{}t_{n}\:
  \widehat{B}(t_{n})\cdots{}\widehat{B}(t_{1})\,\left[q+\frac{t}{m}p\right].
\end{equation} \\

In detail
\begin{align}
  \widehat{B}(t_{1})\left[q+\frac{t}{m}p\right] &= \left(\frac{t-t_{1}}{m}\right)
  \widetilde{F}\!\left[q+\frac{t_{1}}{m}p\right]
  \label{Eqn2.17} \\
  \widehat{B}(t_{2})\widehat{B}(t_{1})\left[q+\frac{t}{m}p\right] &= \left(
  \frac{t_{1}-t_{2}}{m}\right)\left(\frac{t-t_{1}}{m}\right)
  \widetilde{F}\!\left[q+\frac{t_{2}}{m}p\right]\widetilde{F}'\!\left[q
  +\frac{t_{1}}{m}p\right]
  \label{Eqn2.18}
\end{align}
and so on\@. \     Upon letting
\begin{equation}\label{Eqn2.19}
  q+\frac{t_{i}}{m}p = u_{i}
\end{equation}
we obtain, after a systematic analysis, the following general
explicit rule~:
\begin{align}
  \widehat{B}(t_{n}) & \cdots{}\widehat{B}(t_{1})\left[q+\frac{t}{m}p\right] \nonumber \\
  &= \left(
  \frac{t-t_{1}}{m}\right)\widetilde{F}[u_{n}]
  \nonumber \\
  &\quad{} \times\sum\left\{\left(\frac{t_{1}-t_{2}}{m}\right)^{\!\delta(k_{1},2)}\cdots
  \left(\frac{t_{1}-t_{n}}{m}\right)^{\!\delta(k_{1},n)}
  \widetilde{F}^{(k_{1})}[u_{1}]\right.
  \nonumber \\
  &\qquad\qquad{} \times\left(\frac{t_{2}-t_{3}}{m}\right)^{\!\delta(k_{2},3)}
  \cdots\left(\frac{t_{2}-t_{n}}{m}\right)^{\!\delta(k_{2},n)}
  \widetilde{F}^{(k_{2})}[u_{2}]\cdots  \nonumber \\
  &\qquad\qquad{} \times{}\left.
  \left(\frac{t_{n-1}-t_{n}}{m}\right)^{\!\delta(k_{n-1},n)}
  \widetilde{F}^{(k_{n-1})}[u_{n-1}]\right\}
  \label{Eqn2.20}
\end{align}
where
\begin{equation}\label{Eqn2.21}
  \widetilde{F}^{(a)}[u] = \left(\frac{\uD}{\uD{}u}\right)^{\!a}\widetilde{F}[u]
\end{equation}
and the sum in (\ref{Eqn2.20}) is over all $k$'s and $\delta$'s
such that
\begin{equation}\label{Eqn2.22}
  \left. \begin{aligned} k_{1}+k_{2}+\cdots+k_{n-1} = n-1 \\
  \begin{aligned}
    {} & {} \\
    k_{1} &= 1,\ldots,n-1 \\
    k_{2} &= 0,1,\ldots,n-2 \\
    &\;\;\vdots \\
    k_{n-1} &= 0,1
  \end{aligned}\end{aligned}
  \quad\right\}
\end{equation}
and
\begin{align}
  \delta(k_{i},j) &= 0,\quad{}\textrm{if}\quad{}k_{i}=0
  \label{Eqn2.23} \\
  \delta(k_{i},j) &= 0,\quad{}\textrm{if}\quad{}1\leqslant{}j\leqslant{}i
  \label{Eqn2.24}
\end{align}
and for $j>i$, the $\delta(k_{i},j)$ are zero or one such that
\begin{equation}\label{Eqn2.25}
  \sum\limits_{j=i+1}^{n}\delta(k_{i},j) = k_{i},\qquad{}
  i=1,\ldots,n-1
\end{equation}
and
\begin{equation}\label{Eqn2.26}
  \sum\limits_{i=1}^{n-1}\delta(k_{i},j) = 1,\quad{}
  \left(\delta(k_{i},j)=0,\quad{}j=1,2,\ldots,i\Big.\right)
\end{equation}
that is, for a fixed $j$ , $t_{j}$ appears only once in the
product~:
\begin{equation}\label{Eqn2.27}
  \prod\limits_{j=2}^{n}(t_{1}-t_{j})^{\delta(k_{1},j)}
  \prod\limits_{j=3}^{n}(t_{2}-t_{j})^{\delta(k_{2},j)}\cdots.
\end{equation} \\

For example, for $n=4$,
\begin{equation}\label{Eqn2.28}
  k_{1}=1,2,3;\quad{} k_{2}=0,1,2;\quad{} k_{3}=0,1;\quad{}
  k_{1}+k_{2}+k_{3}=3
\end{equation}
and for $k_{1}=2$, $k_{2}=1$, $k_{3}=0$,
\begin{align}
  \delta(2,2)+\delta(2,3)+\delta(2,4) &= 2 \label{Eqn2.29} \\
  \delta(1,3)+\delta(1,4) &= 1 \label{Eqn2.30}
\end{align}
and
\begin{equation}\label{Eqn2.31}
  \delta(2,j)+\delta(1,j) = 1.
\end{equation}
For the harmonic oscillator $V(q)=q^{2}/2$, $\lambda=m\omega^{2}$,
and
\begin{equation}\label{Eqn2.32}
  \widetilde{F}[q]=-q,\quad{} \widehat{B}=-q\frac{\partial}{\partial{}p},\quad{}
  \widehat{A}=\frac{p}{m}\frac{\partial}{\partial{}q}.
\end{equation}
The only solution being
\begin{equation}\label{Eqn2.33}
  k_{1} = \ldots{} = k_{n-1} = 1
\end{equation}
with
\begin{equation}\label{Eqn2.34}
  \widetilde{F}^{(1)}[u_{i}]=-1,\quad{} i=1,\ldots,n-1;\qquad{}
  \widetilde{F}[u_{n}]=-\left[q+\frac{t_{n}}{m}p\right]
\end{equation}
we obtain
\begin{align}
  q(t) =& \sum\limits_{n=0}^{\infty}\left(-\omega^{2}\right)^{n}
  \int_{0}^{t}\!\!\uD{}t_{1}\int_{0}^{t_{1}}\!\!\uD{}t_{2}\cdots{}
  \int_{0}^{t_{n-1}}\!\!\uD{}t_{n}\:(t-t_{1})
  \nonumber \\
  &\quad{} \times{} (t_{1}-t_{2})\cdots{}(t_{n-1}-t_{n})
  \left[q+\frac{t_{n}}{m}p\right]
  \label{Eqn2.35}
\end{align}
which integrates out to
\begin{equation}\label{Eqn2.36}
  q(t) = q\sum\limits_{n=0}^{\infty}\left(-\omega^{2}\right)^{n}
  \frac{t^{2n}}{(2n)!}+\frac{p}{m}
  \sum\limits_{n=0}^{\infty}\left(-\omega^{2}\right)^{n}
  \frac{t^{2n+1}}{(2n+1)!}
\end{equation}
or
\begin{equation}\label{Eqn2.37}
  q(t) = q\cos\omega{}t+\frac{p}{m}\frac{\sin\omega{}t}{\omega}.
\end{equation} \\

\section{Complexification of the Time Evolution}\label{Sec3}

We consider the complex dynamical variable
\begin{equation}\label{Eqn3.1}
  Z(t) = aq(t)+\uI{}bp(t)
\end{equation}
where $a$ and $b$ are arbitrary real constants\@. \     Equation
(\ref{Eqn2.9}) then immediately leads to
\begin{align}
  Z(t) =& \sum\limits_{n=0}^{\infty}\lambda^{n}\int_{0}^{t}\!\!\uD{}t_{1}
  \int_{0}^{t_{1}}\!\!\uD{}t_{2}\cdots\int_{0}^{t_{n-1}}\!\!\uD{}t_{n}\:
  \widehat{B}(t_{n})\widehat{B}(t_{n-1})\cdots{}
  \nonumber \\
  &\quad{} \times{} \widehat{B}(t_{1})\,\left[a\left(q+\frac{t}{m}p\right)+\uI{}bp\right]
  \label{Eqn3.2}
\end{align}
in which
$\widehat{B}(t_{n})\cdots{}\widehat{B}(t_{1})\left[a(q+tp/m)+\uI{}bp\big.\right]$
coincides with the expression in (\ref{Eqn2.20}) except that the
overall factor $(t-t_{1})/m$ in the latter is replaced by
\begin{equation}\label{Eqn3.3}
  \left[a\left(\frac{t-t_{1}}{m}\right)+\uI{}b\right].
\end{equation} \\

A more explicit expression may be also obtained directly from
(\ref{Eqn2.3})\@. \    To this end, consider the Hamiltonian
\begin{equation}\label{Eqn3.4}
  H = \frac{p^{2}}{2m}+m\omega^{2}\frac{q^{2}}{2}+\lambda{}V(q)
\end{equation}
and upon setting
\begin{equation}\label{Eqn3.5}
  Z(t) = q(t)+\uI{}\frac{p(t)}{m\omega}
\end{equation}
one readily obtains
\begin{equation}\label{Eqn3.6}
  Z(t) = \exp\left[-\uI\frac{t\omega}{2}\widehat{C}\right]Z(0)
\end{equation}
where
\begin{align}
  \widehat{C} &= (Z-Z^{*})\left(\frac{\partial}{\partial{}Z}
  +\frac{\partial}{\partial{}Z^{*}}\right)
  \nonumber \\
  &\quad{} +\left\{(Z+Z^{*})-\frac{2\lambda}{m\omega^{2}}
  \widetilde{F}\!\left[\frac{(Z+Z^{*})}{2}\right]\Bigg.\right\}
  \left(\frac{\partial}{\partial{}Z}-\frac{\partial}{\partial{}Z^{*}}\right)
  \label{Eqn3.7}
\end{align}
where $\widetilde{F}[\ldots]$ is defined in (\ref{Eqn2.13})\@. \
For $V(q)=-q$,
\begin{equation}\label{Eqn3.8}
  \widehat{C} = \left[2\left(Z-\frac{\lambda}{m\omega^{2}}\right)
  \frac{\partial}{\partial{}Z}-2\left(Z^{*}-\frac{\lambda}{m\omega^{2}}\right)
  \frac{\partial}{\partial{}Z^{*}}\Bigg.\right]
\end{equation}
leading from
\begin{equation}\label{Eqn3.9}
  Z(t) = \exp\left[-\uI{}t\omega\left(Z-\frac{\lambda}{m\omega^{2}}\right)
  \frac{\partial}{\partial{}Z}\Bigg.\right]Z
\end{equation}
and upon using the identity
\begin{equation}\label{Eqn3.10}
  \exp\left[aZ\frac{\partial}{\partial{}Z}\right]Z = Z\exp{}a
\end{equation}
for any constant $a$, to the expression
\begin{equation}\label{Eqn3.11}
  Z(t) = \frac{\lambda}{m\omega^{2}}+\uE^{-\uI{}t\omega}
  \left(Z-\frac{\lambda}{m\omega^{2}}\right).
\end{equation} \\

A more interesting application is to the geometrical phase
associated with the famous Foucault pendulum with Hamiltonian
\begin{equation}\label{Eqn3.12}
  H = \frac{p_{x}^{2}}{2m}+\frac{p_{y}^{2}}{2m}+\frac{mg}{2L}\left(x^{2}+y^{2}\big.\right)
  +\left(p_{x}y-p_{y}x\big.\right)\omega_{z}
\end{equation}
where $\sqrt{g/L}\equiv\omega_{0}\gg\omega$,
$\omega_{z}=\omega\sin\lambda$, with $\lambda$ denoting the
latitude and $L$ the length of the pendulum\@. \     Then
\begin{align}
  \widehat{O} &= \left(\frac{p_{x}}{m}+y\omega_{z}\right)\frac{\partial}{\partial{}x}
  -\left(m\omega_{0}^{2}x-p_{y}\omega_{z}\big.\right)\frac{\partial}{\partial{}p_{x}}
  \nonumber \\
  &\quad{} +\left(\frac{p_{y}}{m}-x\omega_{z}\right)\frac{\partial}{\partial{}y}
  -\left(m\omega_{0}^{2}y+p_{x}\omega_{z}\big.\right)\frac{\partial}{\partial{}p_{y}}.
  \label{Eqn3.13}
\end{align} \\

Upon defining
\begin{align}
  U &= \left(x+\frac{\uI{}p_{x}}{m}\right)+\uI\left(y+\frac{\uI{}p_{y}}{m}\right)
  \label{Eqn3.14} \\
  V &= \left(x+\frac{\uI{}p_{x}}{m}\right)-\uI\left(y+\frac{\uI{}p_{y}}{m}\right)
  \label{Eqn3.15}
\end{align}
then
\begin{equation}\label{Eqn3.16}
  Z = x+\uI{}y = \frac{U+V^{*}}{2}
\end{equation}
and
\begin{align}
  \widehat{O} &= \left(\frac{\omega_{0}+\omega_{z}}{\uI}\right)
  U\frac{\partial}{\partial{}U}-\left(\frac{\omega_{0}+\omega_{z}}{\uI}\right)
  U^{*}\frac{\partial}{\partial{}U^{*}}
  \nonumber \\
  &\quad{} +\left(\frac{\omega_{0}-\omega_{z}}{\uI}\right)
  V\frac{\partial}{\partial{}V}-\left(\frac{\omega_{0}-\omega_{z}}{\uI}\right)
  V^{*}\frac{\partial}{\partial{}V^{*}}
  \label{Eqn3.17}
\end{align}
which lead to
\begin{equation}\label{Eqn3.18}
  Z(t) = \exp\left[-\uI{}t\left(\omega_{0}+\omega_{z}\big.\right)U
  \frac{\partial}{\partial{}U}\right]
  \exp\left[\uI{}t\left(\omega_{0}-\omega_{z}\big.\right)V^{*}
  \frac{\partial}{\partial{}V^{*}}\right]
  \left(\frac{U+V^{*}}{2}\right).
\end{equation}
With initial conditions
\begin{equation}\label{Eqn3.19}
  x(0) = \uE^{\uI{}2\pi}x_{0},\quad{} y(0) = 0,\quad{}
  \dot{x}(0) = 0,\quad{} \dot{y}(0) = 0
\end{equation}
where $\uE^{\uI{}2\pi}$ denotes the initial phase of the plane of
oscillations of the pendulum moving clockwise, we have
\begin{equation}\label{Eqn3.20}
  U(0) = \uE^{\uI{}2\pi}x_{0} = V^{*}(0)
\end{equation}
giving
\begin{equation}\label{Eqn3.21}
  Z(t) = \uE^{-\uI{}t\omega_{z}}\uE^{\uI{}2\pi}x_{0}\cos\omega_{0}t.
\end{equation}
For $t=2\pi/\omega$, corresponding to a full rotation of the
earth,
\begin{equation}\label{Eqn3.22}
  Z\!\left(\frac{2\pi}{\omega}\right) = \exp\left[\uI{}2\pi\left(1
  -\sin\lambda\big.\right)\Big.\right]
  x_{0}\cos\left(2\pi\frac{\omega_{0}}{\omega}\bigg.\right)
\end{equation}
giving the familiar geometrical Berry phase
$\exp\left[\uI{}2\pi(1-\sin\lambda)\big.\right]$, which is
independent of $g$ and $\omega$, with $2\pi(1-\sin\lambda)$
denoting the solid angle subtended at
the center of the earth\@. \\

\section{A Path Integral Expression}\label{Sec4}

Upon integrating Hamilton's equations
$\partial{}H/\partial{}p=\dot{q}$,
$\partial{}H/\partial{}q=-\dot{p}$ we have
\begin{equation}\label{Eqn4.1}
  f[q(t),p(t)] = f\!\!\left[q(0)+\int_{0}^{t}\!\!\uD{}\tau\:
  \frac{\partial{}H(\tau)}{\partial{}p(\tau)}\;,\;p(0)-\int_{0}^{t}\!\!\uD{}\tau\:
  \frac{\partial{}H(\tau)}{\partial{}q(\tau)}\right].
\end{equation}
We devide the time interval from $0$ to $t$ into $n$ subintervals
each of length $t/n$\@. \     Evaluating each of the summands at
the left-end points of these subintervals, we may rewrite the
right-hand side of (\ref{Eqn4.1}) as
\begin{equation}\label{Eqn4.2}
  \lim\limits_{n\to\infty} f\!\!\left[q_{0}+\frac{t}{n}\sum\limits_{k=0}^{n-1}
  \frac{\partial{}H_{k}}{\partial{}p_{k}}\;,\;p_{0}-\frac{t}{n}\sum\limits_{k=0}^{n-1}
  \frac{\partial{}H_{k}}{\partial{}q_{k}}\right]
\end{equation}
where now we set $q(0)=q_{0}$, $p(0)=p_{0}$, and $k$ is a
short-hand for $tk/n$\@. \     The expression (\ref{Eqn4.2}) to
which the limit $n\to\infty$ is to be taken may be equivalently
rewritten as
\begin{align}
  &\int\!\!\uD{}q_{n}\uD{}p_{n}\:\delta\!\left(q_{0}+\frac{t}{n}\sum\limits_{k=0}^{n-1}
  \frac{\partial}{\partial{}p_{k}}H_{k}-q_{n}\right)\:\delta\!\left(p_{0}
  -\frac{t}{n}\sum\limits_{k=0}^{n-1}\frac{\partial}{\partial{}q_{k}}H_{k}
  -p_{n}\right)f[q_{n},p_{n}]
  \nonumber \\
  &\qquad\qquad{}= \int\left[\Bigg.\prod\limits_{k=1}^{n}\uD{}q_{k}\uD{}p_{k}\:
  \delta\!\left(q_{k-1}+\frac{t}{n}\frac{\partial}{\partial{}p_{k-1}}H_{k-1}
  -q_{k}\right)\right.
  \nonumber \\
  &\qquad\qquad\qquad\qquad{} \left.\times\delta\!\left(p_{k-1}
  -\frac{t}{n}\frac{\partial}{\partial{}q_{k-1}}H_{k-1}
  -p_{k}\right)\Bigg.\right]f[q_{n},p_{n}]
  \nonumber \\
  &\qquad\qquad{}= \int\left[\Bigg.\prod\limits_{k=1}^{n}\uD{}q_{k}\uD{}p_{k}\:
  \delta\!\left(\frac{t}{n}\left\{\frac{\partial}{\partial{}p_{k-1}}H_{k-1}
  -\frac{(q_{k}-q_{k-1})}{t/n}\right\}\right)\right.
  \nonumber \\
  &\qquad\qquad\qquad\qquad{} \left.\times\delta\!\left(
  \frac{t}{n}\left\{\frac{\partial}{\partial{}q_{k-1}}H_{k-1}
  +\frac{(p_{k}-p_{k-1})}{t/n}\right\}\right)\Bigg.\right]f[q_{n},p_{n}].
  \label{Eqn4.3}
\end{align} \\

Upon taking the limit $n\to\infty$ of (\ref{Eqn4.3}),
Eq.~(\ref{Eqn4.1}) becomes
\begin{equation}\label{Eqn4.4}
  f[q(t),p(t)] = \int\!\mathscr{D}[q]\,\mathscr{D}[p]\;
  \delta\!\left(\frac{\partial{}H}{\partial{}p}-\dot{q}\right)
  \delta\!\left(\frac{\partial{}H}{\partial{}q}+\dot{p}\right)
  f[q(t),p(t)]
\end{equation}
as a resolution of the identity as applied to arbitrary functions
$f[q(t),p(t)]$ of $q(t)$ and $p(t)$\@. \     The integrals are
over all paths in phase space, from $0$ to $t$, starting from
$\left(q(0),p(0)\big.\right)$, and the delta functionals as
obtained directly from Hamilton's equations restrict these paths
to the classic one obeying Hamilton's equations at each instant of
time\@. \     Finally, we have also used the fact that
$q_{n}\to{}q(t)$, $p_{n}\to{}p(t)$ from the very definitions in
(\ref{Eqn4.2}), (\ref{Eqn4.3}) for $n\to\infty$\@. \     Equation
(\ref{Eqn4.4}) is in the spirit of the resolution of the identity
of a self-adjoint operator in quantum physics as applied to any
arbitrary vector in the underlying Hilbert space\@. \\

For applications, (\ref{Eqn4.4}) may be rewritten in the more
manageable form
\begin{align}
  \int\!\mathscr{D}[q]\,\mathscr{D}[p]\,\mathscr{D}[\lambda]\,\mathscr{D}[\eta]\;
  \exp\left[\Bigg.\uI\int_{0}^{t}\!\!\uD{}\tau{}\right.
  & \left\{\lambda(\tau)\left(\frac{\partial{}H(\tau)}{\partial{}p(\tau)}
  -\dot{q}(\tau)\right)\right.
  \nonumber \\
  &\quad{} \left.\left.+\eta(\tau)\left(\frac{\partial{}H(\tau)}{\partial{}q(\tau)}
  +\dot{p}(\tau)\right)\right\}\Bigg.\right] f[q(t),p(t)]
  \label{Eqn4.5}
\end{align}
in terms an uncountable infinite number of Lagrange multipliers
$\lambda(\Cdot)$, $\eta(\Cdot)$\@. \\

For example, for a charged particle of charge $e$ in a uniform
magnetic field $B$, say, along the $z$-axis
$\left[\vec{B}=\left(0,0,B\big.\right)\big.\right]$, we may write
for the vector potential
$\vec{A}=\left(-q_{2},q_{1}\big.\right)B/2$, with motion in a
plane, and for the Hamiltonian $H$
\begin{equation}\label{Eqn4.6}
  H = \frac{1}{2m}\left(p_{1}+q_{2}\frac{m\omega}{2}\right)^{\!2}
  +\frac{1}{2m}\left(p_{2}-q_{1}\frac{m\omega}{2}\right)^{\!2}
\end{equation}
with $\omega\equiv{}eB/mc$, $\lambda=(\lambda_{1},\lambda_{2})$,
$\eta=(\eta_{1},\eta_{2})$\@. \     The time-integrand in the
exponential in (\ref{Eqn4.5}), without the $\uI$ factor, may be
rewritten as
\begin{align}
  & \lambda_{1}\left[\frac{1}{m}\left(p_{1}+q_{2}\frac{m\omega}{2}\right)
  -\dot{q}_{1}\right]+\eta_{1}\left[-\frac{\omega}{2}
  \left(p_{2}-q_{1}\frac{m\omega}{2}\right)+\dot{p}_{1}\bigg.\right]
  \nonumber \\
  &+ \lambda_{2}\left[\frac{1}{m}\left(p_{2}-q_{1}\frac{m\omega}{2}\right)
  -\dot{q}_{2}\right]+\eta_{2}\left[\frac{\omega}{2}
  \left(p_{1}+q_{2}\frac{m\omega}{2}\right)+\dot{p}_{2}\bigg.\right]
  \nonumber \\
  &\qquad\qquad{}\equiv{} -\frac{\lambda_{1}}{2m\omega}\left[
  \left(\frac{\uD}{\uD\tau}+\uI\omega\right)U
  +\left(\frac{\uD}{\uD\tau}-\uI\omega\right)U^{*}
  +\frac{\uD}{\uD\tau}(V+V^{*})\Bigg.\right]
  \nonumber \\
  &\qquad\qquad\qquad{} +\frac{\eta_{1}}{4\uI}\left[
  \left(\frac{\uD}{\uD\tau}+\uI\omega\right)U
  -\left(\frac{\uD}{\uD\tau}-\uI\omega\right)U^{*}
  +\frac{\uD}{\uD\tau}(V-V^{*})\Bigg.\right]
  \nonumber \\
  &\qquad\qquad\qquad{} -\frac{\lambda_{2}}{2m\omega\uI}\left[
  \left(\frac{\uD}{\uD\tau}+\uI\omega\right)U
  -\left(\frac{\uD}{\uD\tau}-\uI\omega\right)U^{*}
  -\frac{\uD}{\uD\tau}(V-V^{*})\Bigg.\right]
  \nonumber \\
  &\qquad\qquad\qquad{} -\frac{\eta_{2}}{4}\left[
  \left(\frac{\uD}{\uD\tau}+\uI\omega\right)U
  +\left(\frac{\uD}{\uD\tau}-\uI\omega\right)U^{*}
  -\frac{\uD}{\uD\tau}(V+V^{*})\Bigg.\right]
  \label{Eqn4.7}
\end{align}
where
\begin{align}
  U &= \left(q_{1}\frac{m\omega}{2}-p_{2}\right)
  +\uI\left(p_{1}+q_{2}\frac{m\omega}{2}\right)
  \label{Eqn4.8} \\
  V &= \left(q_{1}\frac{m\omega}{2}+p_{2}\right)
  +\uI\left(p_{1}-q_{2}\frac{m\omega}{2}\right).
  \label{Eqn4.9}
\end{align}
The coefficients of $\lambda_{1}$, $\eta_{1}$, $\lambda_{2}$,
$\eta_{2}$ on the right-hand sides of (\ref{Eqn4.7}) are all
reals\@. \     Upon integration over $\lambda_{1}$, $\eta_{1}$,
$\lambda_{2}$, $\eta_{2}$, we learn that the real and imaginary
parts of
\begin{equation*}
  \left(\frac{\uD}{\uD{}t}+\uI\omega\right)U\pm{}\frac{\uD}{\uD{}t}V
\end{equation*}
must vanish\@. \     That is, $V(t)=V(0)$, and
$\left(\uD/\uD{}t+\uI\omega\big.\right)U=0$ or
$U(t)=U(0)\,\uE^{-\uI\omega{}t}$\@. \     Upon equating the real
and imaginary parts of each of the latter two equations we obtain
the solution
\begin{align}
  q_{1}(t) &= \left(\frac{q_{1}(0)}{2}+\frac{p_{2}(0)}{m\omega}\right)
  +\left(\frac{q_{1}(0)}{2}-\frac{p_{2}(0)}{m\omega}\right)\cos\omega{}t
  \nonumber \\
  &\qquad{} +\left(\frac{q_{2}(0)}{2}+\frac{p_{1}(0)}{m\omega}\right)\sin\omega{}t
  \label{Eqn4.10} \\
  q_{2}(t) &= \left(\frac{q_{2}(0)}{2}-\frac{p_{1}(0)}{m\omega}\right)
  +\left(\frac{q_{2}(0)}{2}+\frac{p_{1}(0)}{m\omega}\right)\cos\omega{}t
  \nonumber \\
  &\qquad{} +\left(-\frac{q_{1}(0)}{2}+\frac{p_{2}(0)}{m\omega}\right)\sin\omega{}t.
  \label{Eqn4.11}
\end{align} \\

To check the consistency of (\ref{Eqn4.4}), we note that in
reference to the first equality on the right-hand side of
(\ref{Eqn4.3}), that
\begin{equation*}
  \exp\left[\frac{t}{n}\left(\frac{\partial{}H_{k-1}}{\partial{}p_{k-1}}\right)
  \frac{\partial}{\partial{}q_{k-1}}\right]
\end{equation*}
is not quite a translation operator for a function of $q_{k-1}$,
since $\partial{}H_{k-1}/\partial{}p_{k-1}$ may, in general,
depend on $q_{k-1}$ as well\@. \      However, in view of the fact
that the limit $n\to\infty$ is to be taken, this operator may be
indeed taken to have such a property for the accuracy needed\@. \
A similar comment applies to the
\begin{equation*}
  \exp\left[-\frac{t}{n}\left(\frac{\partial{}H_{k-1}}{\partial{}q_{k-1}}\right)
  \frac{\partial}{\partial{}p_{k-1}}\right]
\end{equation*}
operator\@. \     To the accuracy needed
(\ref{Eqn4.1})--(\ref{Eqn4.3}) lead to
\begin{align}
  f[q(t),p(t)] = \lim\limits_{n\to\infty}\int\!\! & \left(\:\prod\limits_{k=1}^{n}
  \uD{}q_{k}\uD{}p_{k}\left\{
  \exp\left[\frac{t}{n}\left(\frac{\partial{}H_{k-1}}{\partial{}p_{k-1}}\right)
  \frac{\partial}{\partial{}q_{k-1}}\right]
  \Bigg.\right.\Bigg.\right.
  \nonumber \\
  &\qquad\qquad\qquad\;{} \times
  \exp\left[-\frac{t}{n}\left(\frac{\partial{}H_{k-1}}{\partial{}q_{k-1}}\right)
  \frac{\partial}{\partial{}p_{k-1}}\right]
  \nonumber \\
  &\qquad\qquad\qquad\;{} \left.\left.\times
  \delta(q_{k-1}-q_{k})\delta(p_{k-1}-p_{k})\Bigg.\right\}
  \Bigg.\right)f[q_{n},p_{n}].
  \label{Eqn4.12}
\end{align}
Upon integration over $\left(q_{k},p_{k}\big.\right)$ with the aid
of the delta functions which eventually pick up the
$\left(q_{0},p_{0}\big.\right)$ value for the former, we obtain
the Poisson-bracket solution
\begin{align}
  f[q(t),p(t)] &= \lim\limits_{n\to\infty}\left(
  \exp\left[\frac{t}{n}\frac{\partial{}H(0)}{\partial{}p(0)}
  \frac{\partial}{\partial{}q_{0}}\right]\;
  \exp\left[-\frac{t}{n}\frac{\partial{}H(0)}{\partial{}q(0)}
  \frac{\partial}{\partial{}p_{0}}\right]\Bigg.\right)^{\!n}f[q_{0},p_{0}]
  \nonumber \\
  &= \exp\left[t\left(\frac{\partial{}H(0)}{\partial{}p(0)}
  \frac{\partial}{\partial{}q_{0}}-\frac{\partial{}H(0)}{\partial{}q(0)}
  \frac{\partial}{\partial{}p_{0}}\right)\Bigg.\right]f[q_{0},p_{0}].
  \label{Eqn4.12}
\end{align} \\

\section*{Acknowledgment}

This work was supported by the ``Royal Golden Jubilee Ph.D.
Program'' by the Thailand Research Fund (Grant No. PHD/0022/2545)\@. \\


\begin{thebibliography}{99}
\raggedright

\bibitem{Abrikosov_1993}
A.~A.~Abrikosov~Jr. (1993): ``Path Integral in Constrained
Classical Mechanics'', \textit{Phys.~Lett.}~\textbf{A182} (2--3),
pp.~179--183.

\bibitem{Berry_1984}
M.~V.~Berry (1984): ``Quantal Phase Factors Accompanying Adiabatic
Changes'', \textit{Proc.~Roy.~Soc.~(London)}~\textbf{A392},
pp.~45--57.

\bibitem{Gozzi_1989}
E.~Gozzi, M.~Reuter and W.~D.~Thacker (1989): ``Hidden BRS
Invariance in Classical Mechanics. II'',
\textit{Phys.~Rev.}~\textbf{D40} (10), pp.~3363--3377.

\bibitem{Schwartz_1977}
C.~Schwartz (1977): ``A Classical Perturbation Theory'',
\textit{J.~Math.~Phys.}~\textbf{18} (1), pp.~110--112.

\bibitem{Shapere_1989}
A.~Shapere and F.~Wilczek eds. (1989), \textit{Geometric Phases in
Physics}, World Scientific, Singapore.

\bibitem{Wetterich_1997}
C.~Wetterich (1997): ``Quantum Dynamics in Classical Time
Evolution of Correlation Functions'',
\textit{Phys.~Lett.}~\textbf{B399} (1--2), pp.~123--129. \newline
[ArXiv:~hep-th/9702125]

\end{thebibliography}
\end{document}